\documentclass[osajnl,twocolumn,showpacs,superscriptaddress,10pt]{revtex4-1} 
\usepackage{amsmath,amssymb,graphicx}
\begin{document}

\title{Characteristic functions Describing the Power Absorption 
Response of Periodic Structures to Partially Coherent Fields}

\author{Christophe Craeye}\email{Corresponding author: christophe.craeye@uclouvain.be}
\affiliation{Universit\'e catholique de Louvain, ICTEAM Institute
Place du Levant 3, 1348 Louvain-la-Neuve, Belgium}

\author{Stafford Withington}
\author{Christopher N. Thomas}
\affiliation{Cambridge University, Cavendish Laboratory,
J.J. Thomson Avenue, CB3 OHE Cambridge, UK}

\begin{abstract}
{Many new types of sensing or imaging surfaces are based on periodic thin films.
It is explained how the response of those surfaces to partially coherent fields can be 
fully characterized by a set of functions in the wavenumber spectrum domain. 
The theory is developed here for the case of 2D absorbers with TE illumination and
arbitrary material properties in the plane of the problem, except for the resistivity
which is assumed isotropic.
Sum and differerence coordinates in both spatial and spectral domains 
are conveniently used to represent the characteristic functions, which
are specialized here to the case of periodic structures.
Those functions can be either computed or obtained experimentally.
Simulations rely on solvers based on periodic-boundary conditions,
while experiments correspond to Energy Absorption Interferometry (EAI),
already described in the literature.
We derive rules for the convergence of the representation
versus the number of characteristic functions used, as well as
for the sampling to be considered in EAI experiments. 
Numerical examples are given for the case of 
absorbing strips printed on a semi-infinite substrate.}
\end{abstract}


\maketitle 

\section{Introduction}

Periodic absorbing structures are ubiquitous in detector systems at sub-mm and far-infrared wavelengths.  
At the lowest level, the thin-film absorbers of individual bolometric detectors may be patterned on sub-wavelength 
scales to realise a frequency selective surface \cite{perera2006optical}, or to allow the removal of substrate heat 
capacity while still matching the impedance of the film to free space \cite{mauskopf1997composite}.  At the highest level, 
an imaging array of such detectors can itself be modelled as a periodic absorbing surface.  Periodic structures may 
also be used for wavelength filtering \cite{bossard2006design}. The radiation of interest at these wavelengths is 
often thermal in origin, for example in passive imaging for astronomy, earth observation and security screening.  
In this case the radiation can no longer be assumed spatially coherent over the whole of 
the absorbing surface, as it arises from a collection of incoherent emitters.  An understanding of how partially 
coherent fields interact with periodic surfaces is therefore critical to understanding and optimizing the behaviour 
of these systems. 

This paper provides a framework for the characterisation of the absorption 
response of such periodic surfaces to fields in any state of spatial coherence.
The correlation between fields will be written with the help of
a finite series of eigenfunctions $\phi_n$, as described in \cite{man95}.
Powers absorbed for each of those eigenfunctions will then simply add up.
For general surfaces, those powers will be written in wavenumber
spectral domain as a reaction between field eigenfunctions and
a cross-spectral power density $P^\circ$. Sum and difference coordinates will be
used in both spatial and spectral domains. For periodic surfaces $P^\circ$ will
be shown to be a discrete spectrum (numbered with index $v$) versus spectral difference coordinates, 
while its dependence on spectral sum coordinates is fully described by a set
of characteristic functions ${\cal H}_v$.

Those characteristic functions can be obtained either through measurement or simulation.
An appropriate measurement technique is Energy Absorption Interferometry
(EAI), first mentioned in \cite{sak07} and \cite{wit07}, simulated in 
in \cite{tho10} and put in practice \cite{tho12}. Briefly, a pair of sources is used to 
illuminate the surface and the absorbed power is recorded as the relative
phase between sources is rotated. We explain how the
correlation functions $C_{12}$ obtained from those experiments can be
exploited to compute characteristic functions ${\cal H}_v$. This theory may be viewed
as an extension of \cite{sak07} in the following respects: $(i)$
introduction of the evanescent part of the spectrum, such that near-field
sources or scatterers can be included in the analysis, $(ii)$ use of sum-and-difference
spectral coordinates, $(iii)$ precise link with EAI data and $(iv)$ specialization
to periodic structures, while general surfaces as well as laterally homogeneous
surfaces will also be treated. Numerical examples will be given for the simple
case of strips printed onto a semi-infinite medium. The numerical simulation of EAI experiments
involves the response of the periodic structure to a single source; as introduced
in \cite{wit11}, such a response can be obtained from the periodic-source
case with the help of the Array Scanning Method (ASM, \cite{mun79}).
A phasor notation will be used throughout the paper, with an $\exp (j\,\omega\,t)$ time 
dependence (with $\omega$ radian frequency and $\sqrt{-1}=j$) suppressed.

The remainder of this paper is organized as follows. Section 2 provides
a general spectral-domain representation of power absorbed by an arbritrary surface.
Section 3 specializes that result to 1D-periodic surfaces through the
definitions of characteristic functions ${\cal H}_v$ of spectral sum coordinates.
Section 4 explains how the cross-spectral power density $P^\circ$ can be obtained
from EAI experiments over arbitrary surfaces, while Section 5 specializes
this result for periodic structures and provides a simple rule regarding the 
convergence of the proposed representation versus number of characteristic functions. 
Section 6 makes use of the ASM to explain how the characteristic functions can be
obtained with periodic boundary condition solvers. Section 7 summarizes
the results for the special case of laterally invariant surfaces (but
still arbitrary versus depth). Section 8 provides numerical results 
for periodic surfaces made of strips printed on a semi-infinite medium.
A summary and conclusions are provided in Section 9. To ease the
reading of the paper, the main quantities defined in this paper are summarized
in Appendix B.

\section{Power absorbed by an arbitrary 2D {surface}}
\label{sec:gen}

\begin{figure}[htb]
\begin{center}
\includegraphics[scale=0.8]{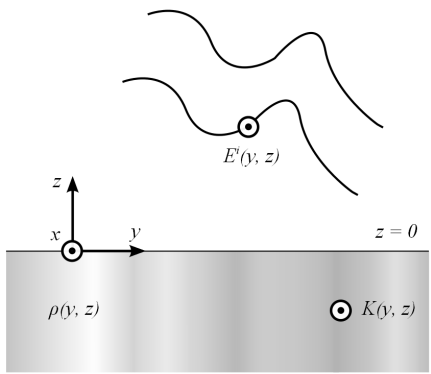}
\caption{Partially coherent field incident on arbitrary semi-infinite medium,
leading to induced current density $K(x,y)$. \label{fig:inc}}
\end{center}
\end{figure}

In this section, we provide a general expression for the power absorbed by a 
2D semi-infinite absorber, based on the response of the absorber in terms of 
induced conduction currents (see sketch in Fig.~\ref{fig:inc}). The excitations correspond 
to incident plane waves, which can be either propagating or evanescent. The result makes 
use of previously published {coherent-mode} 
representations of {partially-coherent} fields \cite{man95}.

The absorbing material occupies a semi-infinite space defined by $z<0$ and is  invariant along $x$. 
It is characterized by its real permittivity, its equivalent 
conductivity (including effect of dielectric losses)
 and its real permeability, with arbitrary dependence on $y$ and $z$ coordinates.  
The inverse of the equivalent conductivity will be denoted by $\rho(y,z)$ and the induced conduction 
current density by $K(y,z)$. Incident fields are assumed to have TE polarization; the extension to 
TM polarization should be straightforward.  Likewise, the extensions to doubly periodic structures and 
to anisotropic resistivity should be straightforward as well.

 Let us denote by $E^i(y)$ the incident field 
at the $z=0$ level; it can be expressed  in spectral domain as $\tilde{E}^i(k_y)$:
\begin{equation}
E^i(y) = \frac{1}{2\,\pi}\ \int_{-\infty}^\infty\, \tilde{E}^i(k_y)\, \exp(-j\,k_y\,y)\, dk_y.
\end{equation}
The current density induced by the field in the absorber may then be expressed in the form:
\begin{equation}
K(y,z) = \frac{1}{2\,\pi}\ \int_{-\infty}^\infty\, \tilde{E}^i(k_y)\, K^\circ(y,z|k_y)\, dk_y
\end{equation}
where $K^\circ(y,z|k_y)$ is the current density in 
response to a unit-magnitude incident field at the $z=0$ level,
with $\exp(-j\,k_y\,y)$ horizontal dependence. 
The expectance value for absorbed power per unit length along $x$ is:
\begin{equation}
P = \frac{1}{2}\, \int_{-\infty}^0\, \int_{-\infty}^\infty\, \rho(y,z)\, 
\left\langle|K(y,z)|^2\right\rangle\,dy\,dz \label{eq:dissip}
\end{equation}
where angle brackets denote ensemble average. Using (1) and (2), the power can be rewritten as
\begin{equation}
P = \frac{1}{2}\,\frac{1}{(2\,\pi)^2}\, \int_{k_{y1}}\int_{k_{y2}}\ 
\tilde{\mathcal{E}}^i(k_{y1}, k_{y2})
\, P^\circ(k_{y1},k_{y2})\, dk_{y1}\,dk_{y2} \label{eq:p00} 
\end{equation}
{where} 
\begin{equation}
\tilde{\mathcal{E}}^i(k_{y1}, k_{y2}) = 
\langle \tilde{E}^i(k_{y1})  \tilde{E}^{i,\star}(k_{y2}) \rangle
\end{equation} 
{is the spectral cross-correlation function of the incident field and}
\begin{equation}
P^\circ(k_{y1},k_{y2}) = \int_{S}\,\rho\, K^\circ(y,z|k_{y1})\,K^{\circ,\star}(y,z|k_{y2})\, dS
\label{eq:pcirc}
\end{equation}
where S refers to the half space $z<0$.  $P^\circ$
is a response function that fully characterises the absorption behaviour of the surface.  
Hereafter $P^\circ$ will be named the {\it cross-spectral power density}.
Here ``spectral'' refers to the spatial wavenumber spectrum and ``cross-spectral''
refers to the fact that it depends on two spectral coordinates.
In the above expressions, the angle brackets denote ensemble averages.
We have assumed that the field is a stationary random process, such that the different 
frequency components are incoherent.

Equation (\ref{eq:p00}) {shows that the absorbed power is a function of the second-order 
spatial correlations in the incident field, as characterised by $\tilde{\mathcal{E}}^i(k_{y1}, k_{y2})$.  
In the case of a Gaussian random field (as is the case for most thermal fields), 
knowledge of the second-order correlation function allows all higher-order moments to be calculated 
\cite{man95}.  A conceptually appealing representation of a partially-coherent field is given in terms 
of its coherent modes \cite{wol86}.  These form a set of functions 
$\tilde{\phi}_i(k_{y})$ that diagonalise the correlation function,} 
\begin{equation}
\tilde{\mathcal{E}}^i(k_{y1}, k_{y2}) = \sum_n \gamma_n
\tilde{\phi}_n (k_{y1}) \tilde{\phi}^\star_n (k_{y2}),
\label{eq:phi}
\end{equation}
so that partial coherence in the field can be thought of as arising from mutually 
incoherent superposition of these individually fully coherent fields.
Substituting this representation into (\ref{eq:p00}) {we obtain the result}
\begin{eqnarray}
P &=& \frac{1}{8\,\pi^2}\,\sum_n\, \gamma_n\, \int_{k_{y1}}\int_{k_{y2}}\ 
\tilde{\phi}_n(k_{y1}) \, P^\circ(k_{y1},k_{y2}) \nonumber \\
&& \hspace{4cm} \tilde{\phi}_n^\star(k_{y2}) dk_{y1}\,dk_{y2},  \label{eq:p0}
\end{eqnarray}
{use of which will be made later}.


This formulation may be viewed as an extension of 
equation (56) of \cite{sak07} to the evanescent part of the spectrum,
which allows the analysis of the response of the sensor to near-field sources.
Contrary to (56) of \cite{sak07}, the analysis presented here
is kept in frequency domain.

\section{Power absorbed by a {periodic 2D surface}}

\begin{figure}[htb]
\begin{center}
\includegraphics[scale=0.8]{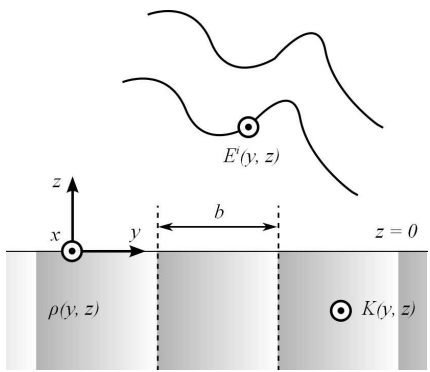}
\caption{Partially coherent field incident on semi-infinite medium
with period $b$ along $y$ coordinate. \label{fig:per}}
\end{center}
\end{figure}

\begin{figure}[htb]
\begin{center}
\includegraphics[scale=0.8]{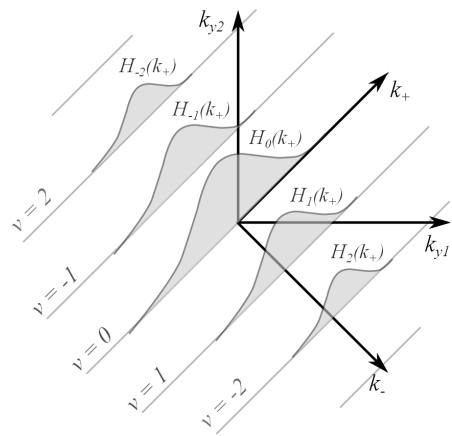}
\caption{Structure of cross-spectral power density in sum-and-difference
spectral coordinates: discrete spectrum characterized by ${\cal H}_v(k^+)$ functions.
\label{fig:struc}}
\end{center}
\end{figure}

Let us now assume that the absorbing structure is periodic with period $b$ along $y$, while 
the structure of the incident wave remains entirely arbitrary (see sketch in Fig.~\ref{fig:per}). 
The integrand appearing in the expression, (\ref{eq:pcirc}), for the cross-spectral power density involves
three factors. The first one, the resistivity $\rho(y,z)$, is periodic versus $y$
with period $b$, while the second and third ones, current densities $K^\circ(y,z|k_{y1})$ 
and $K^{\circ,\star}(y,z|k_{y2})$, are also periodic with same period but with linear phase 
progressions at rates $k_{y1}$ and $-k_{y2}$, respectively. Hence, the product between
those three functions is periodic along $y$ with period $b$, except for a phase progression
at rate $k_{y1} - k_{y2}$. Hence, the integrand of (\ref{eq:pcirc}) corresponds, 
after multiplication by a phase factor $\exp (j\,(k_{y1}-k_{y2})\,y)$, to a purely periodic function. 
Therefore, we can write the Fourier series:
\begin{eqnarray}
&& \rho(y, z) \, K^\circ(y,z|k_{y1})\,K^{\circ,\star}(y,z|k_{y2})  =  \nonumber \\
&& \hspace*{0.2cm} \frac{1}{2\pi} e^{-j\,(k_{y1}-k_{y2})\,y}\,
\sum_{v=-\infty}^\infty\, {Q}_v(z|k_{y1},k_{y2})\, e^{j\,v\,y\,2 \pi/b}
\label{eq:pper1}
\end{eqnarray}
with
\begin{eqnarray}
&& {Q}_v(z|k_{y1},k_{y2}) = \nonumber\\
&& \frac{2\,\pi}{b}\ \int_{0}^{b}
\rho(y, z) \, K^\circ(y,z|k_{y1})\,K^{\circ,\star}(y,z|k_{y2})\, \nonumber \\
&& e^{j\,(k_{y1}-k_{y2})\,y}\, e^{-j\,v\,y\,2 \pi/b}\ dy.
\label{eq:pper2}
\end{eqnarray}
Then, introducing that Fourier series into 
(\ref{eq:pcirc}), rearranging the order of integrations along $y$ and noting that 
$\int_{-\infty}^\infty\, \exp(j\,\zeta)\, d\zeta = 2 \pi\, \delta(\zeta)$, we obtain:
\begin{equation}
P^\circ(k_{y1},k_{y2}) = \sum_{v=-\infty}^\infty\, {\cal H}_v(k^+)\, \delta(k^- - v\,\frac{2 \pi}{b})
\label{eq:p0cal}
\end{equation}
with the following definitions of sum and difference wavenumbers:
\begin{equation}
k^+ = (k_{y1} + k_{y2})/2\ \ \ \ \mbox{and}\ \ \ \ k^- = k_{y1} - k_{y2} \label{eq:sumdif}
\end{equation}
and  the following characteristic function
\begin{eqnarray}
{\cal H}_v(k^+) &=& \frac{2 \pi}{b}\, \int_{S_\circ}\, \rho\, 
K^\circ\left(y,z|k^+ + v\, \frac{\pi}{b}\right) \nonumber \\
&& K^{\circ,\star}\left(y,z|k^+ - v\, \frac{\pi}{b}\right)\, dS_\circ 
\label{eq:calh}
\end{eqnarray}
where $S_o$ is the unit cell of the periodic structure ($0 \leq y<b$).  
We have also used the $k^- = 2 \pi v / b$ determination imposed by the presence
of the delta function to write $k_{y1}$ 
and $k_{y2}$ in terms of $k^+$. The induced currrent density $K^\circ$ over the unit cell, 
and hence also the characteristic functions ${\cal H}_v$ can be computed with any type 
of periodic boundary conditions solver. The structure of the cross-spectral power
density $P^\circ$ is sketched in Fig.~\ref{fig:struc}.

Inserting (\ref{eq:p0cal}) into (\ref{eq:p0}), the definition of the 
characteristic function ${\cal H}_v(k^+)$ yields the following general expression for 
the power absorbed into periodic structures of period $b$:
\begin{eqnarray}
P &=& \frac{1}{8\,\pi^2}\,\sum_n\, \gamma_n\, \sum_{v=-\infty}^\infty\, \int_{k^+}\, 
\tilde{\phi}_n \left(k^+ + v\,\frac{\pi}{b}\right)  \nonumber \\
&& \hspace*{2cm}  \tilde{\phi}_n^\star\left(k^+ - v\, \frac{\pi}{b}\right)\,
\, {\cal H}_v(k^+)\,dk^+.
\label{eq:pp}
\end{eqnarray}
In the following two sections, we propose a methodology for obtaining the ${\cal H}_v$ functions from data obtained 
through {an interferometric measurement technique}.

\section{Characterization through Energy Absorption Interferometry}
\label{sec:eai}

\begin{figure}[htb]
\begin{center}
\includegraphics[scale=0.8]{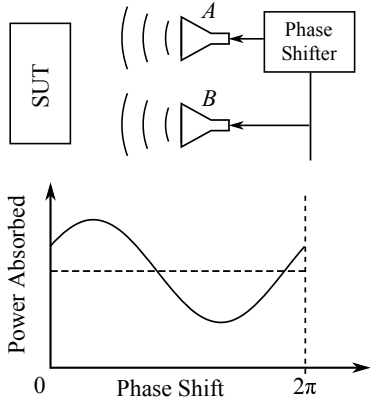}
\caption{Basic operating principle of Energy Absoption Interferometry.
\label{fig:eai}}
\end{center}
\end{figure}

{Energy Absorption Interferometry (EAI) is a technique that was originally proposed for detector 
characterisation \cite{wit07}, and has recently been demonstrated 
experimentally in this context \cite{tho12}.  More recently, its use as a general 
tool for characterising power absorption in structures has been studied \cite{with12}.  
As illustrated in Fig.~\ref{fig:eai}, EAI involves illuminating the Structure Under Test (SUT) with a pair of 
phase-locked sources and recording the power absorbed.  
As the relative phase between two sources is rotated, 
the detector output displays a fringe pattern, the complex amplitude of which characterises the 
detector response. This section explains how the response function $P^\circ$ defined in} (\ref{eq:pcirc}) 
{can be measured from this fringe amplitude, allowing the calculation of absorbed power for 
any incident field. This will be done first for arbitrary (i.e. non-periodic) absorbers.}

{In the 2D case treated here the sources are infinite line sources oriented along $x$ and of intensity $I$}.  
The dissipated power can be written as:
\begin{equation}
P^l = \frac{1}{2}\,\int\, \rho\, |K^l|^2\ dS
\end{equation}
where $K^l$ corresponds to the current density excited by a line
source of intensity $I$ located a height $z=h$ and horizontal coordinate
$y=y_s$. Superscript $l$ reminds us that the absorbed power has been obtained
with the line-source illumination in an EAI experiment.

It is easy to prove that, for two sources, with intensities $I_1 = I$ and $I_2 = I\,e^{j\,\phi}$, 
located at $(y_{s} = y_1,h)$ and  $(y_{s}=y_2,h)$ respectively, we have:
\begin{eqnarray}
P^l(y_1,y_2) = \frac{1}{2}\, \left( \alpha_{11} + 
\alpha_{22} + 2\,\alpha_{12}\, \cos (\phi - \beta_{12}) \right)
\label{eq:nine}
\end{eqnarray}
with correlation functions defined as:
\begin{eqnarray}
C_{ij}(y_1,y_2) &=& \alpha_{ij}\,e^{j\,\beta_{ij}}  \\
&=& \!\!\! \int_{-\infty}^0\!\!\int_{-\infty}^{\infty}\!\! \rho\, K^l_i(y,z) K_j^{l,\star}(y,z)\,dy dz \label{eq:cij2}\\
&=& \int\, \rho\, K_i^l\,K_j^{l,\star}\ dS
\label{eq:cij1}
\end{eqnarray}
where $K^l_i$ is the current density induced by the line source located at $y=y_{si}$. 
Combining experimental results with one and two sources, the correlation function
$C_{12}$ is readily obtained. 
The remainder of this section consists of providing a link between that correlation function and 
the cross-spectral power density $P^\circ(k_{y1},k_{y2})$, defined in (\ref{eq:pcirc}) 
and that completely characterizes the absorber.

We start with the current induced in the absorber when only one line source is 
present, with current $I$ and position $(y,z)=(y_s,h)$. The incident field 
at $z=0$ from the current source is decomposed into plane waves as (\cite{mor53}, Sec. 7.3):
\begin{equation}
E^i(y) = -\frac{I\,k\,\eta}{2\,\pi}\, \int_{-\infty}^\infty
\,\frac{e^{-j\,k_y\,(y-y_s)} e^{-j k_z h}}{2\,k_z}\,  dk_y
\end{equation}
with the following constraint:
\begin{equation}
k_y^2 + k_z^2 = k^2 = \omega^2 / c^2
\label{eq:square}
\end{equation}
where $\omega$ is the radian frequency and $c$ is the speed of light in free space.
The corresponding induced current density $K^l$ is obtained by simply
introducing in the integrand the current density $K^\circ(y,z|k_y)$ obtained
with a unit-magnitude incident electric field with $e^{-j\,k_y\,y}$ dependence:
\begin{equation}
K^l_i(y,z) = - \frac{I\,k\,\eta}{2\,\pi}\ \int_{-\infty}^\infty\,
K^\circ(y,z|k_y)\, 
\frac{e^{j\,k_y\,y_{s,i}} e^{-j k_z h}}{2\,k_z}\ dk_y.
\end{equation}
From this plane-wave decomposition, the correlation (\ref{eq:cij1}) can be linked with
the cross-spectral power density $P^\circ$ defined in (\ref{eq:pcirc}):
\begin{eqnarray}
C_{12}(y_1,y_2) &=& \frac{k^2\,\eta^2\,|I|^2}{4\,(2\,\pi)^2}\,
\int_{k_{y1}}\!\!\int_{k_{y2}}\, P^\circ(k_{y1},k_{y2}) \nonumber \\
&& \hspace*{-1cm} \frac{e^{j\,(k_{y1}\,y_1 - k_{y2}\,y_2)}\, e^{-j\,(k_{z1} \pm k_{z2})\,h}}{k_{z1}\,k^\star_{z2}}\,
dk_{y1}\,dk_{y2}
\end{eqnarray}
where the $+$ sign is taken when $k_{z2}$ becomes imaginary.
Since it will be interesting to see to what extent correlations
depend on average positions, besides dependence on difference between coordinates.
To clarify this, we make the following sum-and-{difference} change of variables:
\begin{equation}
r = (y_1 + y_2)/2, \ \ \ \ \ \ s = y_1 - y_2 \label{eq:rs}
\end{equation}
Together with (\ref{eq:sumdif}) this allows us to rewrite the correlation as:
\begin{eqnarray}
C_{12}(y_1,y_2) &=& \frac{k^2\,\eta^2\,|I|^2}{4\,(2\,\pi)^2}\,
\int_{k^+}\!\!\int_{k^-}\, P^\circ(k_{y1},k_{y2})\, \nonumber\\
&& \hspace*{-0.5cm} \frac{e^{j\,(k^+\,s + k^-\,r)}\, e^{-j\,(k_{z1} \pm k_{z2})\,h}}{k_{z1}\,k^\star_{z2}}\,
dk^+\,dk^-
\label{eq:ft1}
\end{eqnarray}
where $(k_{y1},k_{y2})$ wavenumbers (and related $(k_{z1},k_{z2})$ quantities, see 
(\ref{eq:square})) can be readily obtained from the definitions (\ref{eq:sumdif}) of $k^+$ and $k^-$. 
It is important to notice that equation (\ref{eq:ft1}) has the form of a 2D Fourier 
transform. This observation leads us to the fact that the cross-spectral power density
$P^\circ(k_{y1},k_{y2})$ can be obtained through inverse Fourier transformation:
\begin{eqnarray}
P^\circ(k_{y1},k_{y2})  &=&
\frac{4\,k_{z1}\,k_{z2}^\star}{k^2\,\eta^2\,|I|^2}\,
e^{j\,(k_{z1} \pm k_{z2})\,h} \nonumber \\
&& \hspace*{-0.7cm} \int_r\,\int_s\ C_{12}(r,s)\ e^{-j\,(k^+\,s + k^-\,r)}\ dr\,ds
\label{eq:pcft}
\end{eqnarray}
where sum and difference coordinates have been used in both
spatial and spectral domains. Regarding the latter, in order to
simplify the notation, we have mixed $(k_{y1},k_{y_2})$ coordinates 
(and related $(k_{z1},k_{z_2})$ wavenumbers) with $(k^+,k^-)$
coordinates, with the link between variables given by (\ref{eq:sumdif}). 

\section{Characterization of a {periodic 2D absorber}}
\label{sec:per}

\begin{figure}[htb]
\begin{center}
\includegraphics[scale=0.8]{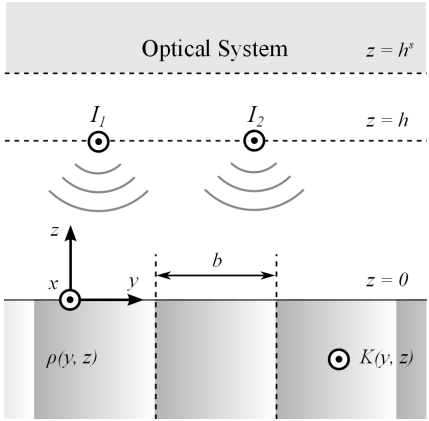}
\caption{Energy Absorption Interferometry above a periodic surface.
\label{fig:eaiper}}
\end{center}
\end{figure}

In this section, the periodicity of the structure is introduced by noticing that 
the correlation function is periodic versus the sum coordinate $r=(y_1+y_2)/2$.  
A Fourier series representation of the correlation function obtained from EAI 
(Fig.~\ref{fig:eaiper}) leads 
to a discrete-spectrum representation of the cross-spectral power density and, in turn, 
to a combined discrete and continuous spectral representation of the absorbed power.
A direct link is established between correlation functions and characteristic functions ${\cal H}_v$
describing the periodic absorber.

If $y_1$ and $y_2$ are the positions of the sources, we may represent the correlations
in the $RS$ plane, defined through (\ref{eq:rs}). In view of the periodicity
of the absorber, we may expect the correlation function to be periodic of period $b$ 
versus sum coordinate $r$. This allows us to write the dependence of $C_{12}$ as a Fourier
series along $r$, in which each term is a function of $s$:
\begin{equation}
C_{12}(y_1,y_2) = C_{12}(r,s) =
\frac{1}{2\,\pi}\, \sum_{v=-\infty}^\infty\, e^{j\,\frac{2\, \pi}{b}\,v\,r}\, H_v(s).
\label{eq:chv}
\end{equation}
The $H_v$ functions can be obtained from the correlation data as:
\begin{equation}
H_v(s) = \frac{2\,\pi}{b}\,\int_0^{b}\, C_{12}(r,s)\, e^{-j\,\frac{2\,\pi}{b}\,v\,r}\ dr.
\label{eq:hvb}
\end{equation}

The cross-spectral power density (\ref{eq:pcft}) can then be re-written for the specific case of
periodic absorbers via the following observation:
\begin{eqnarray}
&&\int_r\,\int_s\ C_{12}(r,s)\ e^{-j\,(k^+\,s + k^-\,r)}\ dr\,ds  = \nonumber\\
&& \hspace*{-0.8cm} 
\frac{1}{2\,\pi}\, \sum_v\, \int_s\, H_v(s)\, e^{-j\,k^+\,s} \int_r\, e^{j\,v\,\frac{2\,\pi}{b}\,r}\,
e^{-j\,k^-\,r}\ dr\,ds
\end{eqnarray}
where the integral over $r$ produces $2\,\pi\, \delta\left(k^- - v\, {2\,\pi}/{b}\right)$.
From there, the following result is obtained for the cross-spectral power density (\ref{eq:pcft}):
\begin{eqnarray}
P^\circ(k_{y1},k_{y2}) &=&
\frac{4\,k_{z1}\,k_{z2}^\star}{k^2\,\eta^2\,|I|^2}\,
e^{j\,(k_{z1} \pm k_{z2})\,h}\, \nonumber \\
&& \hspace*{-1.2cm} \sum_v \int_s\ H_v(s)\ e^{-j\,k^+\,s}\,ds\ \delta\left(k^- - v\, \frac{2\,\pi}{b}\right)
\label{eq:sumint}
\end{eqnarray}

This result allows us to write the absorbed power (\ref{eq:p0}) 
for arbitrary partially-coherent incident fields as:
\begin{eqnarray}
P &=& \frac{1}{2\,\pi^2}\, 
\frac{1}{k^2\,\eta^2\,|I|^2}\,
\sum_n \gamma_n \sum_v\int_{k^+} \,k_{z1}\,k_{z2}^\star\, \nonumber\\
&& e^{j\,(k_{z1} \pm k_{z2})\,h}\,
\tilde{\phi}_n(k_{y1})\, \tilde{\phi}_n^\star(k_{y2}) \nonumber\\
&& \int_s\ H_v(s)\ e^{-j\,k^+\,s}\,ds\ dk^+
\label{eq:Psum}
\end{eqnarray}
where, as a result of (\ref{eq:sumdif}) and of the presence of a Dirac delta function
in (\ref{eq:sumint}), we have
\begin{equation}
k_{y1} = k^+ + v\, \pi / b\ \ \ \ \ \mbox{and}\ \ \ \ \  k_{y2} = k^+ - v\, \pi / b.
\label{eq:ky12}
\end{equation}

To ease further interpretation, in a different notation, the absorbed power reads:
\begin{eqnarray}
P &=& M_0\ 
\sum_n \gamma_n \sum_v\int_{-\infty}^\infty\, k_{z1}\, k_{z2}^\star\, \nonumber\\
&& e^{j\,(k_{z1} \pm k_{z2})\,h}\,
\tilde{\phi}_n(k_{y1})\, \tilde{\phi}_n^\star(k_{y2})\, \tilde{H}_v(k^+)\, dk^+
\label{eq:newp}
\end{eqnarray}
where $M_0 =  1/2\, \left(\pi\, k\, \eta |I|\right)^{-2}$ and $\tilde{H}_v(k^+)$ is the
Fourier transform of $H_v(s)$. For $v=0$, the interpretation is very simple:
for each coherent eigenfield $\phi_i$, the absorbed power is proportional to an
inner product between the spectral density $|\tilde{\phi}_i|^2$ of the incident
field and the sensitivity $\tilde{H}_0$, given as the Fourier transform of the
average correlation (averaging over absolute coordinates $r$ along unit cell $b$, within a 
factor $2\,\pi$) obtained from the EAI experiments. For $v \not= 0$, the interpretation
is more complex, the spectral density is replaced by the $\tilde{\phi}_i(k_{y1})\, \tilde{\phi}_i^\star(k_{y2})$
product and the sensitivity is replaced by the Fourier transform of the $H_v(s)$ function,
corresponding to a non-fundamental component of the periodic (versus $r$) correlation function.

From the expression (\ref{eq:newp}) of absorbed power, it is now clear that the function
$\exp({j\,(k_{z1} \pm k_{z2})\,h})\, \tilde{H}_v(k^+)$ completely defines
the power response of the periodic structure. The function $\tilde{H}_v(k^+)$
depends on the height at which the EAI experiments have been carried out
and the $\exp({j\,(k_{z1} \pm k_{z2})\,h})$ factor propagates those results to zero 
height. Comparing forms (\ref{eq:pp}) and (\ref{eq:newp}) of absorbed power, 
one obtains the following link between characteristic functions of the absorber
and Fourier transform of correlation functions obtained from EAI experiments:
\begin{equation}
{\cal H}_v(k^+) =
e^{j\,(k_{z1} \pm k_{z2})\,h}\, \tilde{H}_v(k^+) \frac{4\,k_{z1}\,k_{z2}^\star}{k^2\,\eta^2\,|I|^2}.
\label{eq:link}
\end{equation}
The minus sign before $k_{z2}$ has to be used when $|k^+ - v\,\pi/b| > k$ (upward
evanescent wave). We here recall the link (\ref{eq:square}) between $k_z$ and $k_y$ and 
the link (\ref{eq:ky12}) between $k_y$ and $k^+$ and $v$. Those relations will be important
for the understanding of the following.

One may wonder about the importance of integrating (\ref{eq:pp}) outside the visible region, 
i.e. for $|k^+| > k$, where evanescent waves are involved. 
It is well known that energy transfer can actually take place via 
evanescent waves, provided that there are transmitted and reflected waves.
Those evanescent waves may be radiated by the observed object located in the
near field or may be scattered by parts of the observation instrument. 
Let us denote by $h^s$ the lowest parts of the observed sources or of the instrument 
(see Fig.~\ref{fig:eaiper}). That height corresponds to the characteristic distance over which
evanescent waves may propagate toward the surface; evanescent waves propagating
over longer distances will simply undergo stronger decay. When either $k_{y1}$ or $k_{y2}$,
as given by (\ref{eq:ky12}), lies outside the visible region ($|k_y| > k$), the function 
$F = \tilde{\phi}_n(k_{y1})\, \tilde{\phi}_n^\star(k_{y2})$
decays by a factor $|\exp( {-j\,(k_{z1} \pm k_{z2})\,h^s})|$ \cite{wol86}.
In the expression (\ref{eq:pp}) for the absorbed power, that decay
combines with the exponential factor appearing in the expression of ${\cal H}_v(k^+)$,
given in (\ref{eq:link}). This leads to a decay ($D>1$) in total power given by:
\begin{equation}
D = |e^{ {-j\,(k_{z1} \pm k_{z2})\,(h^s - h)}}|
\end{equation}
where $k_{z1}$ or $k_{z2}$ (or both) can be imaginary, thus leading to exponential decay.

The convergence of the proposed representation versus index $v$ can be 
estimated in the limit of large $v$. We estimate here a lower bound for $D$. 
Wavenumbers $k_{z1}$ and $k_{z2}$ are linked to $k_{y1}$ and $k_{y2}$ through
(\ref{eq:square}). We also know that $k_{y1} - k_{y2} = v\, 2\pi/b$, through (\ref{eq:ky12}).
The slowest decay rate will be obtained when one of the wavenumbers, say 
$k_{y2}$, lies within the visible region, which imposes $|k_{y2}| < k$.
For large values of $v$, such that $v\, 2\pi/b \gg k $, we 
obtain $k_{z1} \simeq -j\, v\, 2\pi/b$ (from (\ref{eq:square}) for large $k_y$). 
This leads to a very simple model for the slowest decay versus index $v$:
\begin{equation}
D \simeq e^{|v|\,\frac{2\,\pi}{b}\,(h-h_s)}
\end{equation}
which, in dB scale, corresponds to a decay by a factor
\begin{equation}
D_{dB} \simeq \frac{20\,\pi}{\ln 10}\, \frac{h_s-h}{b}\, |v|.
\label{eq:ddb}
\end{equation}
This simple model does not account for the link between incident plane waves and induced 
currents, but this link is material-dependent and there is no a priori known reason
for having a strong (e.g. exponential) dependence on $v$ of induced currents. 
Also, the model proposed above is well supported by the simulation data presented in 
Section \ref{sec:simu}, where that model will be shown to be very accurate.

To avoid ill-posedness in the characterization of the periodic absorber, 
it makes sense to avoid too high values of $D$, which suggests 
$h<h^s$ as a recommended mode of operation if one wants to include the evanescent part of the 
spectrum in the analysis. In other words, if the observed sources or parts of the instrument 
are in the near-field of the detector, it is advised to conduct EAI experiments at even lower height.

The above observations regarding convergence have important implications in terms of spatial 
sampling rate in the EAI experiments. First,  for a given order $v$, function $H_v$ should only be 
estimated if the signal-to-noise ratio exceeds $D_{dB}$, as given in (\ref{eq:ddb}).
Having determined the maximum value of $|v|= V_{max}$, it is known that the interval $b$ 
should be sampled with at least $2 V_{max}$ points. Given the fast convergence of
successive harmonics, it may not be necessary to use many more points:
if all harmonics $|v| > V_{max}$ are truly neglible, then $2 V_{max}$ points is strictly sufficient.

Also, obviously, 
in view of the periodicity of the structure, the displacement of one of the sources can be limited 
to the unit cell.  The other source needs to be shifted over a longer range; the precise range depends 
on the material properties of the absorber and will not be discussed here in detail. 

\section{Simulation with periodic boundary condition solvers}
\label{sec:num}

The numerical simulation of the response of a given periodic absorber in an EAI experiment 
requires a priori the estimation of the induced current $K^l(y,z)$ for one or two line sources placed 
above the periodic structure. Since the structure is virtually infinite (or at least large compared to 
the wavelength), such a simulation requires important computational resources. Here, as introduced
in \cite{wit11}, we will make use 
of the Array Scanning Method (ASM) \cite{mun79} to express the correlation function in terms of 
results obtained when considering that both absorbers and sources are infinitely periodic, with a linear 
phase progression in the source fields. The advantage of this new formulation is that it relies on 
quantities that can be obtained very fast with the help of unit-cell solvers. Those solvers can be based 
on either differential or integral-equation approaches.

Using the ASM \cite{mun79}, the electric current densities $K$ 
in the periodic material induced by an isolated source at $y=y_s$ can be expressed 
as the superposition of  solutions $K^\infty$ with periodic line sources:
\begin{equation}
K^l(y_\circ + n\,b,z) = \frac{1}{2\,\pi}\ \int_0^{2 \pi}\ K^\infty(y_\circ,z|\psi)_{|y_s}
\ e^{-j\,n\,\psi}\ d\psi
\label{eq:asm}
\end{equation}
where $\psi$ is the phase shift between successive unit cells and $y_\circ$ is assumed to be 
always located within the reference unit cell $S_\circ$, i.e. $0 \le y_\circ < b$. 
The index $y_s$ denotes the position of the line source within the unit cell. To alleviate notations,
the index will be omitted below wherever possible. Besides the explanation given in the 
original publication \cite{mun79}, 
based on the Poisson summation formula,
an alternative proof for (\ref{eq:asm}) is given in \cite{cra11}. As explained in
\cite{cap07}, the infinite-array solution 
$K^\infty$
can be written as an infinite spectral
summation (Floquet modes) in which each term corresponds to the effect
of one plane wave; from there it is easy to show that swapping summation and integration 
in (\ref{eq:asm})
is equivalent to integration over the whole spectral axis.

This allows a representation of dissipated power for a single source. Denoting the
unit cell, i.e. $-\infty < z \le 0$ and $0 \le y_\circ < b$, by $S_\circ$ we have:
\begin{eqnarray}
P &=& \frac{1}{2}\, \sum_{n=-\infty}^\infty\, \int_{-\infty}^0\, \int_0^b\, \rho\, 
|K^l(y_\circ+n\,b,z)|^2\,dy_\circ\,dz \\
&=&\! \frac{1}{2\,(2\,\pi)^2}\!\! \sum_{n=-\infty}^\infty\int_{S_\circ}\rho\,
\left|\int_0^{2\pi} K^\infty(y_\circ,z|\psi) e^{-j\,n\,\psi} d\psi \right|^2 \nonumber\\
 && \hspace*{6cm} dS_\circ \label{eq:pda}\\
&=& \frac{1}{4\,\pi}\,
\int_0^{2\,\pi}\, \int_{S_\circ}\,\rho\, |K^\infty(y_\circ,z|\psi)|^2\, dS_\circ\, d\psi
\label{eq:pdb}
\end{eqnarray}
where the transition from (\ref{eq:pda}) to (\ref{eq:pdb}) is proven in Appendix A. 

In order to obtain the power absorbed in the EAI experiment,
the above result is extended to the case of two sources at height $h$ with relative phase shift $\psi$:
\begin{eqnarray}
P &=& \sum_{n=-\infty}^\infty\,\frac{1}{2}\ \int_{S_\circ}\,\rho
\, |K^l_1(y_\circ + n\,b, z) + \nonumber \\ 
&& \hspace*{2cm} K^l_2(y_\circ + n\,b, z)\,e^{j\,\phi}|^2\, dS_\circ \\
&=& \frac{1}{4\,\pi}
\int_0^{2\,\pi} \int_{S_\circ} \rho\, |K_1^\infty(y_\circ + n\,b,z|\psi) + \nonumber\\
&& \hspace*{1.5cm} K_2^\infty(y_\circ + n\,b,z|\psi)\,e^{j\,\phi}|^2\, dS_\circ\, d\psi \\
&=& \frac{1}{2}\, \left( \alpha_{11} + 
\alpha_{22} + 2\,\alpha_{12}\, \cos (\phi - \beta_{12}) \right) \label{eq:ninebis}
\end{eqnarray}
with
\begin{eqnarray}
C_{ij} &=& \alpha_{ij}\,e^{j\,\beta_{ij}} = \frac{1}{2\,\pi}
\int_0^{2\,\pi}\, \int_{S_\circ}\,\rho\,\ K_i^\infty(y_\circ,z|\psi)\, \nonumber\\
&& \hspace*{2cm} K_j^{\infty,\star}(y_\circ,z|\psi)\,dS_\circ\, d\psi.
\label{eq:cinf}
\end{eqnarray}
We can see that the expression (\ref{eq:ninebis}) has the same form as (\ref{eq:nine}), 
but that the correlation functions (\ref{eq:cinf}) can now be obtained through integration 
of products of periodic-source currents $K^\infty$ over the unit cell $S_\circ$. 
This will make the simulation feasible with limited computational resources, 
without simulating fields in huge finite structures.  

\section{Special case of laterally homogeneous surfaces}
\label{sec:hom}

In case the absorbing material is not periodic but is invariant along $y$,
the results provided thus far can be simplified. A summary of those
results is given in this section, considering an arbitrary variation
of permittivity, permeability and resistivity versus coordinate $z$.

In this case, the cross-spectral power density $P^\circ$ contains only the $v=0$
mode:
\begin{equation}
P^\circ(k_{y1},k_{y2}) = {\cal H}_0(k^+)\, \delta(k_{y1}-k_{y2})
\end{equation}
with $k^+ = k_{y1} = k_{y2}$ and
\begin{equation}
{\cal H}_0(k^+) = 2 \pi\, \int_{-\infty}^0\, \rho\, 
|K^\circ(z|k^+)|^2\, dz. 
\label{eq:calh0}
\end{equation}
The absorbed power in case of partially coherent incident 
fields then simplifies as:
\begin{equation}
P = \frac{1}{8\,\pi^2}\,\sum_n\, \gamma_n\, \int_{-\infty}^\infty\ 
|\tilde{\phi}_n(k^+)|^2 \, {\cal H}_0(k^+)\ dk^+
\end{equation}

Since correlations $C_{12}$ do not depend on absolute coordinate $r$,
(\ref{eq:pcft}) can be rewritten as
\begin{eqnarray}
P^\circ(k_{y1},k_{y2}) &=&
\frac{4\,k_{z1}\,k_{z2}^\star}{k^2\,\eta^2\,|I|^2}\,
e^{j\,(k_{z1} \pm k_{z2})\,h} \nonumber \\
&& \hspace*{-0.3cm} \int_s\ C_{12}(s)\ e^{-j\,k^+\,s}\,ds\ \delta(k_{y1}-k_{y2})
\end{eqnarray}
and $C_{12}$ has only one Fourier component:
\begin{equation}
C_{12}(s) = H_0(s) / (2 \pi)
\end{equation}
From the above, it is then possible to provide a simple expression for
the sole characteristic function:
\begin{equation}
{\cal H}_0(k^+) = \frac{8\,\pi\, k_{z1}\,k_{z2}^\star}{k^2\,\eta^2\,|I|^2}\,
e^{j\,(k_{z1} \pm k_{z2})\,h}\, \tilde{C}_{12}(k^+)
\end{equation}
with
\begin{equation}
\tilde{C}_{12}(k^+) = \int_{-\infty}^\infty\ C_{12}(s)\ e^{-j\,k^+\,s}\,ds.
\end{equation}

\begin{figure}[htb]
\begin{center}
\includegraphics[scale=0.20]{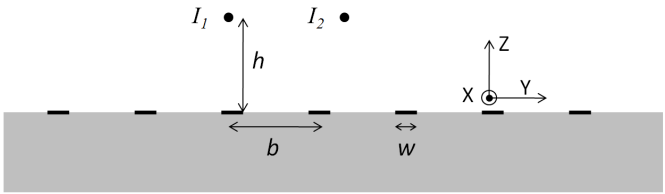}
\caption{General geometry of simulated structure. \label{fig:abs}}
\end{center}
\end{figure}

\section{Simulation examples}
\label{sec:simu}

\begin{figure}[htb]
\begin{center}
\includegraphics[scale=0.45]{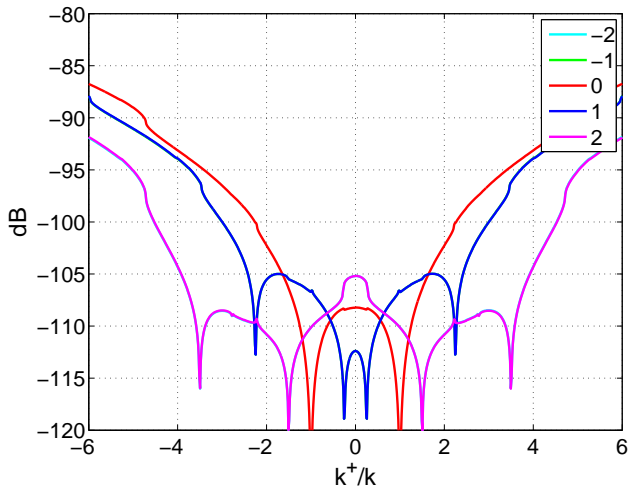}
\caption{Characteristic functions ${\cal H}_v$ for $-2 \le v \le 2$,
 for a height of $h=0.25\,\lambda$ from the surface.\label{fig:hcal}}
\end{center}
\end{figure}
\begin{figure}[htb]
\begin{center}
\includegraphics[scale=0.45]{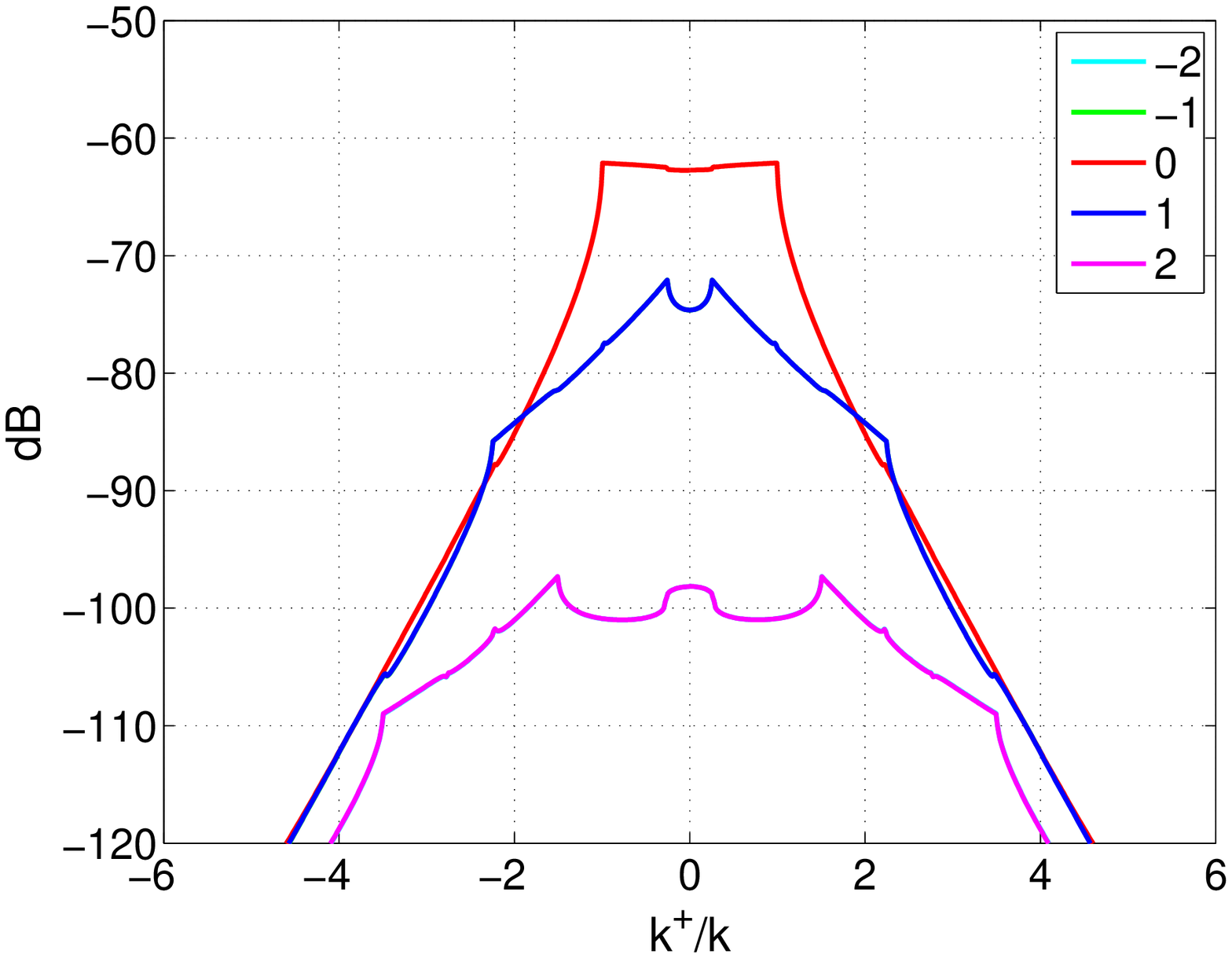}
\caption{$\tilde{H}_v$ functions for a height of $h=0.25\,\lambda$ from the surface.\label{fig:ht}}
\end{center}
\end{figure}
\begin{figure}[htb]
\begin{center}
\includegraphics[scale=0.45]{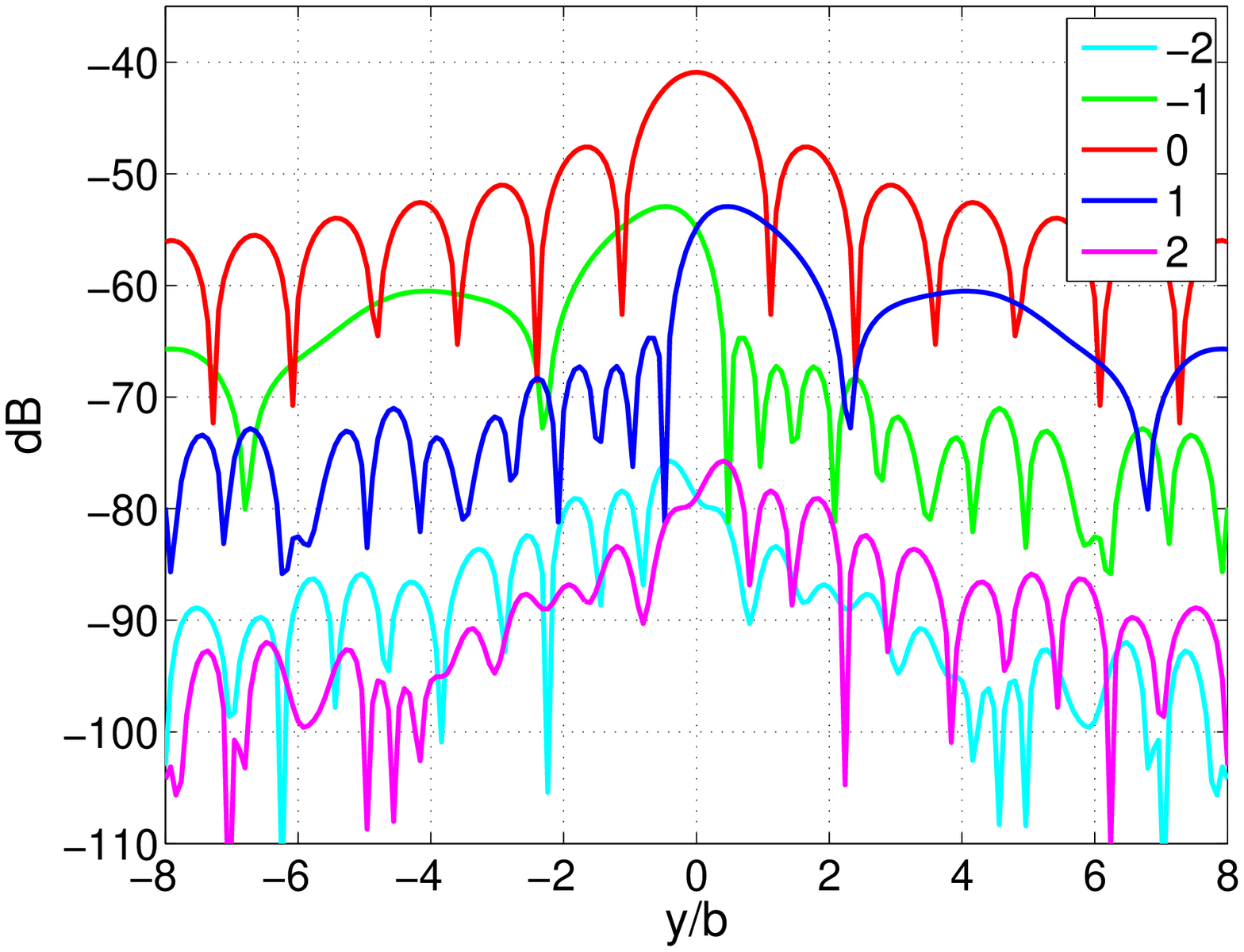}
\caption{$H_v$ functions for a height of $h = 0.25\,\lambda$ from the surface.\label{fig:h}}
\end{center}
\end{figure}
\begin{figure}[htb]
\begin{center}
\includegraphics[scale=0.45]{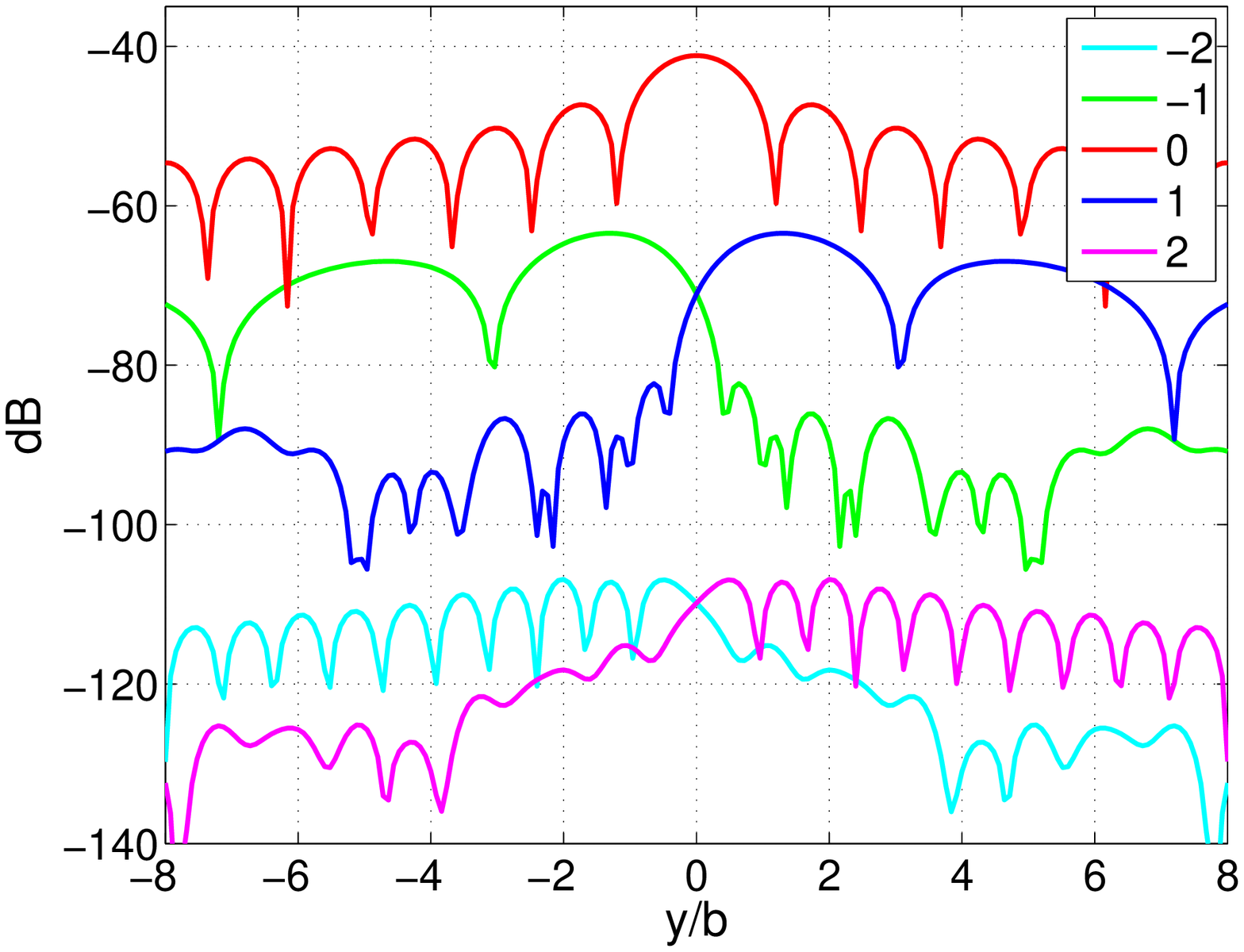}
\caption{${H}_v$ functions for a height of $0.5\,\lambda$ from the surface.\label{fig:h1}}
\end{center}
\end{figure}
\begin{figure}[htb]
\begin{center}
\includegraphics[scale=0.45]{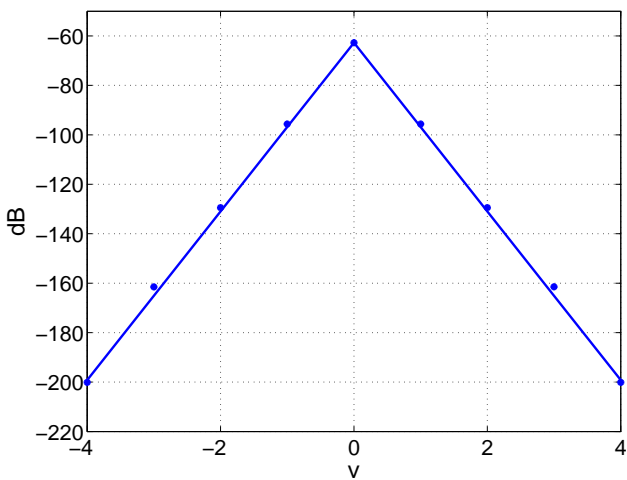}
\caption{Decay of $H_v(0)$ according to simullation (bullets) and according to model (solid), for $h=\lambda/2$.\label{fig:decay}}
\end{center}
\end{figure}

Simulations have been carried out for the case of an absorber made of a grid
of strips placed on a semi-infinite medium (see configuration in Fig.~\ref{fig:abs}).
The periodic method of moments \cite{har93} \cite{cra07} has been used; the Green's function
associated with the semi-infinite medium has been computed in spectral domain \cite{tai71}.
The free-space wavelength $\lambda$ is 1 mm, the spacing $b$ is 0.4 mm,
the width $w$ of the strips is 0.15 mm, the relative permittivity of the medium
in the lower half space is $12 - j\, 0.012$, close to that of undoped silicon,
 and the sheet impedance of the strips is a tenth of 
the free-space impedance, i.e. 37.7 Ohm.  Four basis functions have been used across the width 
of each strip. The characteristic functions ${\cal H}_v(k^+)$ will be shown and commented,
as well as functions $\tilde{H}_v(k^+)$ and ${H_v}(s)$ for two different heights $h$
above the surface.

Fig.~\ref{fig:hcal} shows the magnitudes of ${\cal H}_v$ functions for different values of $v$
for a height above the surface equal to $\lambda/4$. It can be seen that
$|{\cal H}_v| = |{\cal H}_{-v}|$.
It is interesting to note that, in the visible region, i.e. for $|k^+|<k$, the different
functions have the same order of magnitude. Outside the visible region, they
tend to grow rapidly. This raises the question of the possible non-bandlimited
aspect of those functions. However, one should bear in mind that the absorbed
power (\ref{eq:pp}) is computed as a reaction between characterisitic functions
${\cal H}_v$ and Fourier transforms of eigenfields $\phi_n$ estimated at two
different wavenumbers. For large values of the wavenumber $k^+$, eigenfields
are evanescent and hence can compensate the growth of the ${\cal H}_v$ functions.
More precisely, if the minimum height of the source or instrument is $h_s$,
the product $\phi_n()\,\phi_n^\star()$ entails a factor $\exp(j\,(k_{z1} \pm k_{z2})\,h_s)$.
In this expression, $k_{z1}$ and $k_{z2}$ become imaginary outside the visible region, 
which leads to exponential decay. 

For a similar reason, the functions ${\tilde H}_v(k^+)$ also decay for large
values of $k^+$, with a rate proportional to the height $h$ at which EAI
experiments have been carried out. Indeed, the (continuous spectrum of)
plane waves radiated by the line sources are evanescent for large wavenumbers
and hence contribute little to the absorbed power.
This effect is visible in Fig.~\ref{fig:ht}, where the ${\tilde H}_v(k^+)$
functions are shown. This is also evidenced by  
the link (\ref{eq:link}) between characteristic functions ${\cal H}_v$  
and ${\tilde H}_v$ functions obtained from correlation measurements.
Also, as explained in Section \ref{sec:per}, the significance of ${\cal H}_v$
rapidly decreases as $|v|$ increases. The inverse Fourier transforms of those
functions, linked to the EAI experiments according to (\ref{eq:hvb}), are shown in
Fig.~\ref{fig:h}. Here too, it can be seen that the significance of those functions
decays with $|v|$ and $h$. This can be observed from Fig.~\ref{fig:h1}, 
where ${\tilde H}_v$ functions are shown for a larger distance, $h=\lambda/2$.
It can be seen from Figs.~\ref{fig:h} and \ref{fig:h1} that the spatial bandwidth 
of ${H}_v$ functions 
increases with $|v|$, as confirmed by their Fourier transforms, 
visible in Fig.~\ref{fig:ht} for $h = 0.25\,\lambda$.

Finally, the decay of functions $H_v$ versus $v$ is made explicit for $h=0.5 \lambda$
in Fig.~\ref{fig:decay}, where the stars correspond to the result obtained
numerically, while the solid line coresponds to the model (\ref{eq:ddb}).
Since only the decay rate is analyzed here, the value of the model has been
arbitrarily set to the numerically obtained value for $v=0$. Excellent
agreement is observed between modeled and numerically obtained
decay rates.



\section{Summary and perspectives}
\label{sec:conc}

The response of a given absorbing surface to partially coherent fields
can be described with the help of a cross-spectral power density $P^\circ$:
the absorbed power is given as a double spectral integration involving $P^\circ$
and the product of eigenfunctions describing the incoherent field.
It is convenient to represent the spectral integration versus sum and difference
spectral coordinates. In case of laterally homogenous surfaces, $P^\circ$
is a delta function versus the spectral difference coordinate. If the surface
is laterally periodic, $P^\circ$ is a discrete spectrum versus the difference
coordinate, while it is represented by characteristic functions 
${\cal H}_v$ versus the spectral sum coordinate. Those functions correspond
to the integration over the unit cell of the resistivity multiplied by
the product of currents induced by plane waves with different wavenumbers.

Energy Absorption Interferometry (EAI) consists of illuminating the surface
by a pair of phase-locked sources and recording the absorbed power.
We explained how $P^\circ$ can be obtained from the correlation function
produced by such experiments.
In the case of periodic surfaces, one of the sources is swept over
a unit cell, while the other one is swept over a larger domain.
It is recommended to conduct the experiments with sources at height $h$ lower than 
the lowest height $h^s$ of the sources or of the telescope optics.
The observed correlation functions can be decomposed into a Fourier
series versus average positions whose coefficients are functions 
$H_v$ of difference positions. 
We provided a simple link between the Fourier transforms
of the latter functions and the ${\cal H}_v$ functions that characterize 
the periodic absorber. 
From this same representation, we delineated a simple rule regarding 
the convergence of the representation versus the number
of characteristic functions used. In EAI experiments,
the number of samples within a unit cell should be equal 
to at least the number of characteristic functions looked for. 

Finally, with the help of the Array Scanning Method, we explained how
the quantities referred to above can be obtained with the help 
of periodic boundary-condition simulations, while the sources in the
EAI experiments are not periodic. Simulation examples were given
for a very simple structure made of parallel strips printed
on a semi-infinite medium, which allowed us to check the convergence
referred to above.

The main further step in this analysis concerns the extension to doubly
periodic absorbers and to fields with arbitrary polarization. A future
communication will also express the absorber's characteristic functions
in terms of current induced by Floquet waves.

\bibliographystyle{IEEEtran}
\bibliography{eai_periodic_dec13}

\section*{Appendix A: link between (\ref{eq:pdb}) and (\ref{eq:pda})}

Result (\ref{eq:pdb}) can be obtained from
 (\ref{eq:pda}) with the help of the
following identity (dependence on $y$ and $z$ dropped from notation):
\begin{eqnarray}
\sum_{n=-\infty}^\infty\,
\left|\int_0^{2\,\pi}\, K^\infty(\psi)\, e^{-j\,n\,\psi} d\psi \right|^2 &=& \nonumber\\
&& \hspace*{-4cm} \int_0^{2\,\pi}\!\!\int_0^{2\,\pi}  
K^\infty(\psi_1)\, K^{\infty,\star}(\psi_2) \nonumber\\
&& \hspace*{-3cm} \left( \sum_{n=-\infty}^\infty e^{-j\,n\,(\psi_1-\psi_2)}
\right)\ d\psi_1\,d\psi_2
\label{eq:psi2}
\end{eqnarray}
in which the sum between parentheses equals $2\pi\, \sum_{p=-\infty}^\infty\,
\delta(\psi_1 - \psi_2 - p\,2 \pi)$, which allows one to solve one of the integrals in (\ref{eq:psi2})
and where only the $p=0$ term has a non-zero contribution.

\section*{Appendix B: glossary}

The most important quantities defined in this paper are summarized below.
Related equations are between parentheses.
\begin{itemize}
\item{$P$: absorbed power (\ref{eq:dissip}).}
\item{$P^\circ$: cross-spectral power density (\ref{eq:pcirc}).}
\item{$(r,s)$: sum and difference space coordinates (\ref{eq:rs}).}
\item{$(k^+,k^-)$: sum and difference spectral coordinates (\ref{eq:sumdif}).}
\item{$K^\circ(y,z|k_y)$: current density induced by plane wave with wavenumber $k_y$.}
\item{$\phi_n$: eigen-function of partially-coherent incident field (\ref{eq:phi}).}
\item{$C_{12}(r,s)$: correlation function obtained from EAI experiments (\ref{eq:cij1}).}
\item{${\cal H}_v(k^+)$: characteristic function of periodic absorber (\ref{eq:calh}).}
\item{$H_v(s)$: Fourier component of $C_{12}$ versus $r$ (\ref{eq:chv}).}
\item{${\tilde{H}}_v(k^+)$: Fourier transform of $H_v(s)$ (\ref{eq:Psum}).}
\end{itemize}
Two key relationships of this paper are (\ref{eq:pp}) and (\ref{eq:link}), 
which respectively link $P$ with ${\cal H}_v$ and ${\cal H}_v$ with ${\tilde H}_v$.

\end{document}